\documentclass[10pt]{article}
\usepackage{amsmath, amsfonts, amssymb}
\usepackage{epsf}
\begin{document}
\title{Anomalous Pulsars}
\author{I.F.Malov}

\date{}
\maketitle

{Pushchino Radio Astronomy Observatory, P.N.Lebedev
 {Physical Institute, Russian Academy of  Sciences,
  \\ Pushchino, Moscow Region, 142290, Russia\\}

E-mail: {malov@prao.ru}

\begin{abstract}
Many astrophysicists believe that Anomalous X-Ray Pulsars (AXP),
Soft Gamma-Ray Repeaters (SGR), Rotational Radio Transients
(RRAT), Compact Central Objects (CCO) and X-Ray Dim Isolated
Neutron Stars (XDINS) belong to different classes of anomalous
objects with neutron stars as the central bodies inducing all
their observable peculiarities. We have shown earlier (I.F.Malov
and G.Z.Machabeli, Astron. Astrophys. Trans. 25  7, 2006)  that
AXPs and SGRs could be described by the drift model in the
framework of preposition on usual properties of the central
neutron star (rotation periods $P \sim 0.1 - 1$ ~sec, surface
magnetic fields $B \sim 10^{11} - 10^{13}$ ~G). Here we shall try
to show that some differences of considered sources will be
explained by their geometry (particularly, by the angle $\beta$
between their rotation and magnetic axes). If  $\beta  \lesssim
10^{\circ}$ (the aligned rotator) the drift waves at the outer
layers of the neutron star magnetosphere should play a key role in
the observable periodicity. For large values of  $\beta$ (the case
of the nearly orthogonal rotator) an accretion from the
surrounding medium (for example, from the relic disk) can cause
some modulation and transient events in received radiation.

   Key words: AXP, SGR, RRAT, radio pulsars: drift waves, relic disk, accretion
   PACS: 97.60.Gb, 97.10.Kc

\end{abstract}

\section{Introduction}

During the last decades the interest to some X-ray sources has
grown drastically. These are first of all AXPs and SGRs,
"magnetars" as many investigators believe. We showed in our paper
[1] that these objects could be described by the drift model, and
the usual neutron star lied in the base of both types of sources.
The drift waves cause some changes of magnetic field lines
curvature and as the consequence the modulation of emission
propagation directions (Fig.1). The calculations in the framework
of this model have been carried out. As the result some parameters
of neutron stars in the known AXPs and SGRs were obtained: the
interval for the rotation periods is $P = 11 - 737$ ~msec, for
their derivatives $dP/dt = 3.7 \times 10^{-16} - 5.5 \times
10^{-12}$, and for the surface magnetic fields $\log B_s = 11.22 -
13.24$. The main difference between AXPs / SGRs and normal radio
pulsars is a small inclination of the magnetic moment to the
rotation axis in the first group. The corresponding angle $\beta$
must be of order to (or less than) $10^{\circ}$. We believe that
all neutron stars with $\beta  \lesssim 10^{\circ}$ must belong to
the population with observed intervals between successive pulses
(if they are seen) $P_{obs}$ of order to several seconds. For
example, so called high magnetic field pulsars are the objects of
this type. Real rotation periods and magnetic fields for pulsars
with very long observed pulse periods are the same as for normal
pulsars [2]: $P = 0.5 - 1.12$ ~sec, $B_s = (0.2 - 16 ) 10^{12}$
~G. Fig. 2 and 3 show some details of the drift model.

\section{Radio transients}

There were discovered new types of objects during the last dozen
years. These are transients, i.e. sources emitting suddenly a few
number of pulses and then being switched off. The table 1 contains
the list of such sources and possible candidates [3-8].

To explain transient character of emission from these objects some
models were put forward during the last few years.

1. Precession of pulsars with long time nullings [9]. To achieve
the agreement with observations it is necessary to suggest in this
model large angles of precession ($\gamma > 15^{\circ}$). $\gamma$
is the angle between the rotation axis and the angular moment. The
estimations give for normal radio pulsars $\gamma < 10^{\circ}$.
The required precession in neutron stars is possible for very
large deformations only and can be realized at rotation periods of
order to 1 msec. Moreover this precession must quickly damp
(during $10^2 - 10^4$ periods of the precession).  To avoid these
difficulties the authors propose to use as the host object in such
pulsars a solid quark star with possible large elastic
deformations. However the existence of quark stars is rather
problematic, and their corresponding models have not been worked
out up to now.

2. Zhang et al. [10] considered the possibility of the location of
a pulsar near and below the "death line" and irregular appearances
of the pulsar above this line. Such appearances take place when
the magnetic field of the solar spot type comes to the surface and
induces the process of  $e^{\pm}$ - pairs creation. What is the
real equation of the death line is unclear yet. If take the
equation used by the authors some transients find themselves above
this line, and other mechanism of switching off of their radiation
must operate.

\begin{figure}[h]
\epsfysize=10cm \epsfbox{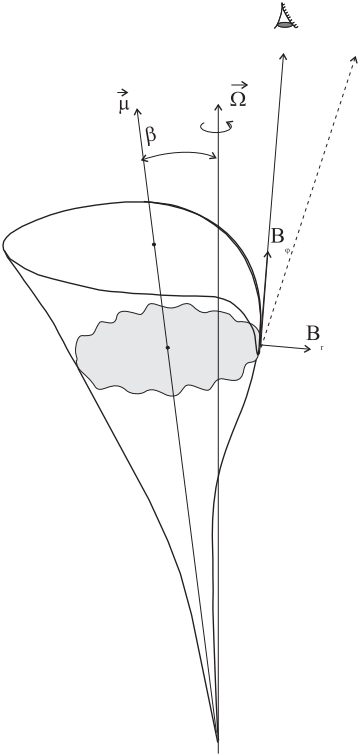}
   \caption{Scheme of the drift model. $\mu$ is the magnetic moment,
                            $\Omega$ is the rotation axis.}
\end{figure}

\begin{figure}[h]
\epsfysize=8cm \epsfbox{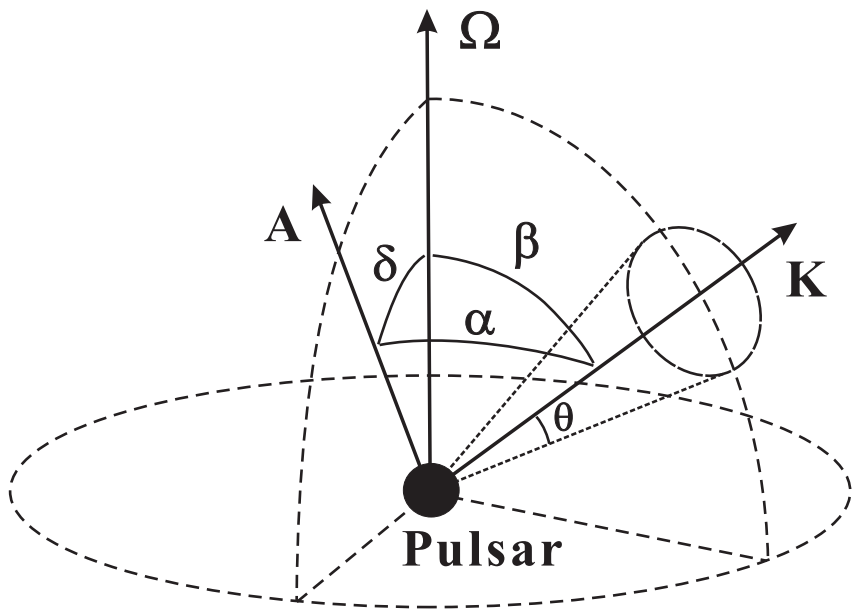}
   \caption{Geometry of the drift model. {\bf K} is the axis of the emission
cone, {\bf A} is the direction to the observer. $\delta =
Constant$, $\theta = Constant$, $\alpha$ and $\beta$ are functions
of time.}
\end{figure}

\begin{figure}[h]
\epsfysize=10cm \epsfbox{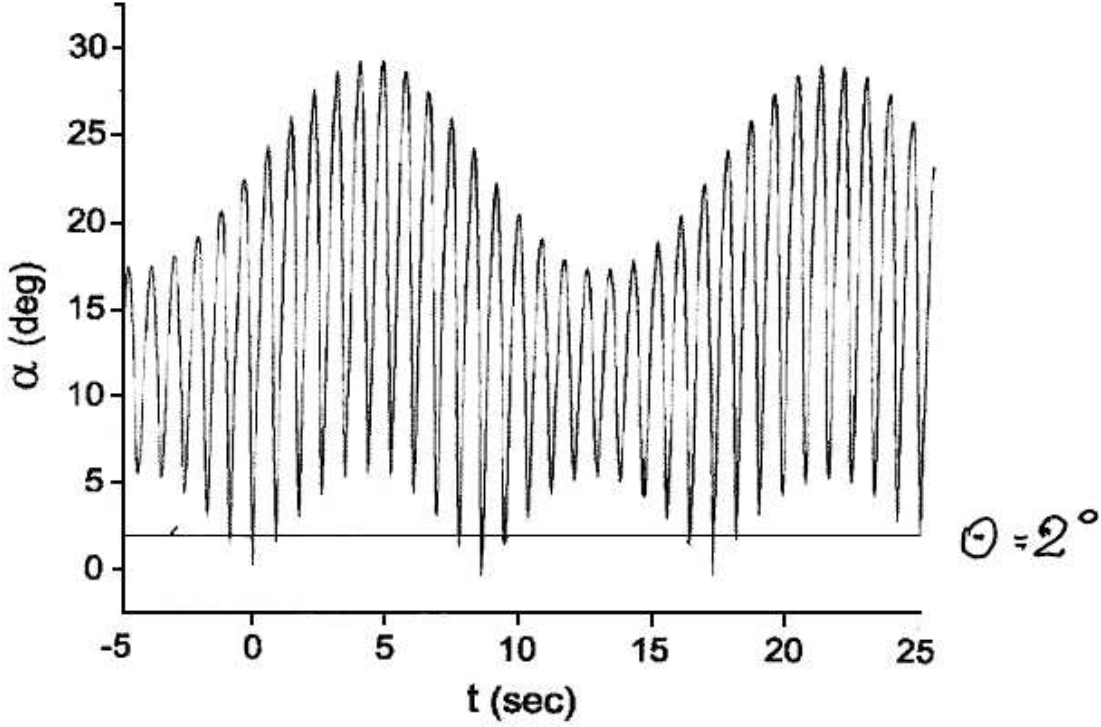}
   \caption{The oscillating behaviour of $\alpha$ with time for $\Omega =2 \pi / 0.85
sec^{-1}, \, \Theta = 2^{\circ}$.}
\end{figure}

\begin{table}
\caption{Radio transients}
\begin{tabular}{ccccccc}
\hline
Name & P & $w_{50}/P$ & $dP/dt$ & B & $\tau$ & $dE/dt$\\
& (sec) & $\%$ &  $10^{-15}$ & $10^{12}$  ~G & $10^6$ years &
$10^{31}$ erg /sec\\ \hline J0848-43     & 5.97748(2)           &
0.50 & - - - - & & & \\ J1317-5759 & 2.6421979742(3) &    0.38
12.6(7)& 5.83(2)& 3.33(2) & 2.69(1)\\ J1443-60     & 4.758565(5) &
0.42 & - - - - & & & \\ J1754-30     & 0.422617(4)         & 3.79
& - - - - & & & \\ J1819-1458 & 4.263159894(6)   &    0.07 576(1)
50.16(6) & 0.1172(3) & 24.94(5)\\ J1826-14     &  0.7706187(3) &
0.26 & - - - - & & & \\ J1832+0031 & & & & & & \\ J1839-01     &
0.93190(1) & 1.61 & - - - - & & & \\ J1846-02 & 4.476739(3) & 0.36
& - - - - & & & \\ J1848-12 & 6.7953(5) & 0.03 & - - - - & & & \\
J1911+00 & & & & & & \\ J1913+1333 & 0.9233885242(1) & 0.22 &
7.87(2) & 2.727(4) & 1.860(6) & 39.4(1)\\ B1931+24 & 0.813690303 &
5 & 7.15-10.79 & 2.6 & 1.6 & $5.9 10^{-6}$ \\ J1649+2533 &
1,0152573918(5) & 2,46 & 0,5594(2) & 0,79 & 28,7 & 2\\ J1752+2359
& 0,409050865044(9) &  0,98 & 0,6427(9) & 0,5 & 10,1 & 39,81\\
B0656+14 & 0.384885 & 7.7 & 55.01 & 4.68 & 0.11 & 72.44\\
J1745-3009 & 4627.8 & & & & & \\\hline
\end{tabular}
\end{table}

\begin{figure}[h]
\epsfysize=10cm \epsfbox{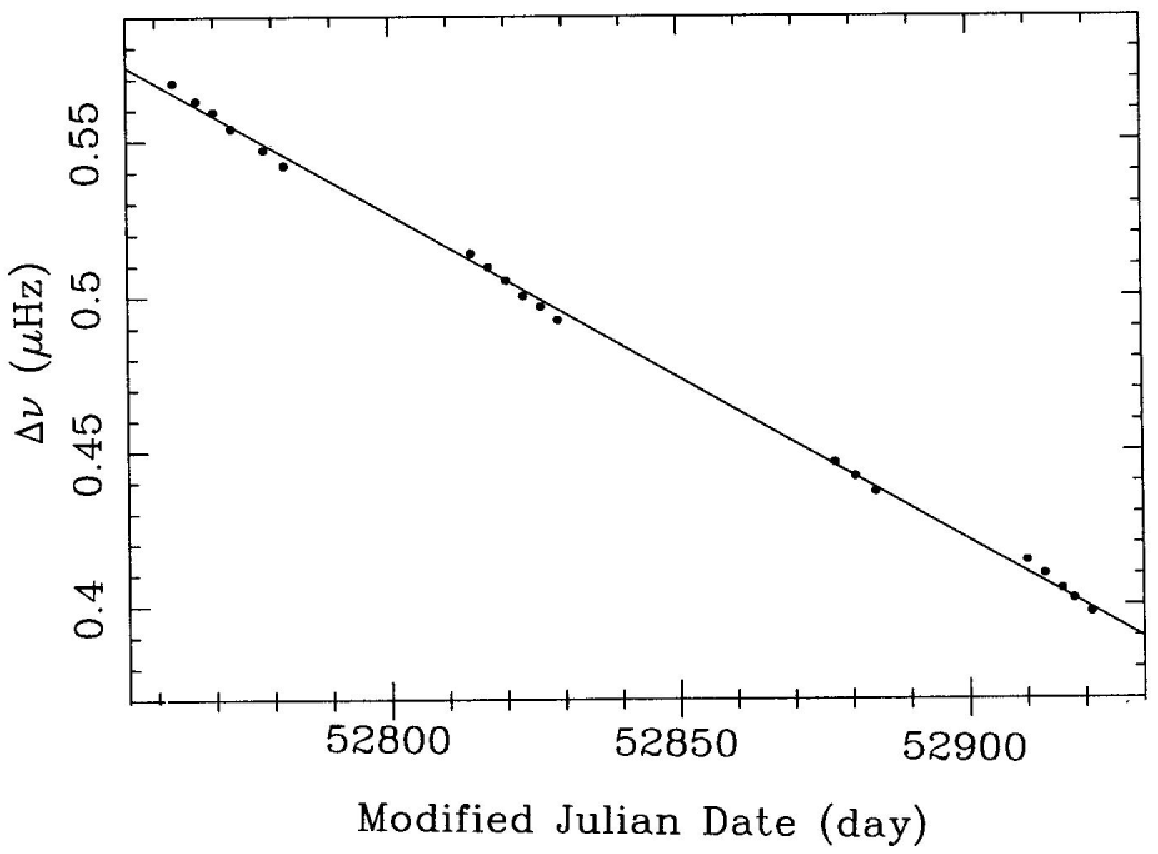}
   \caption{Variation of the rotational frequency of PSR B1931+24
              over a 160-day period. The errors in the measurement of
              the data points are  smaller  than the size of the symbols.
  The best-fit straight-line through the points is shown,
  representing a frequency derivative $d\nu/dt = -12.2 \times 10^{-15}$ Hz
  s$^{-1}$
  (from [4]).}
\end{figure}

3. The same authors [10-11] considered the another possibility to
explain RRATs. They put forward the idea of so called inward
radiation, i.e. radiation directed opposite to usual "right"
pulsar radiation, and connected nullings with this emission
directed inward. In this model the geometric effects play role
only and the breaking must be the same in "on" and "off" phases.
So, some other mechanisms must operate at least in PSR B1931+24,
where the frequency derivatives differ significantly (1.5 times)
in these two phases (Fig.4). Similar behaviour shows PSR
J1832+0031[8] (Fig.5).

4. The long period ($77^m$) of the source J1745-3009 near the
Galactic Center excludes a neutron    star as the central body
because the losses of its rotation energy $dE/dt$ are very low
($\sim 10^{22}$ ~erg/sec) to provide the observable luminosity.
Zhang \& Gil [12] proposed to consider a white dwarf as a pulsar.
It must have a complicated magnetic field at the surface (like the
field of  solar spots) inducing an irregular creation of $e^{\pm}$
-pairs. However there is no known white dwarfs with pulsar
characteristics. Moreover the  rotation period  ($\sim 77^m$) is
of order to minimal values for such objects, and the magnetic
field must be extremely high  ($\sim 10^9$ ~G).

5. Turrola et al.[13] have tried to explain the source J1745-3009
in the framework of the binary system with the orbital period
($77^m$). It was proposed that this system consists of two neutron
stars and is similar to the system of the pulsar J0737-3039. Its
coherent radio emission is generated in the shock wave due to an
interaction between the wind from the more energetic pulsar and
the magnetosphere of its companion. The modulation of emission is
caused by the orbit eccentricity and connected with periodic
penetration of the wind in the inner layers of the companion
magnetosphere. However this model can be realized for
eccentricities $e = 0.4$ and can not explain the lack of emission
in the time interval between the consequence bursts. Moreover both
neutron stars in such a system must be rather powerful pulsars
with periods $0.3-1$ ~sec. The question arises why they are not
observed.  If a pulsar in this system will be detected this model
will be tested by the change of the period with time.

6. Drift model.

In this model [14] the line of sight could find itself sometimes
 inside  of
an emission cone due to  cataclysms at the   surface of the
neutron star. It is suggested that the angle  $\beta$ between the
rotation axis and the magnetic moment is small enough. A very
narrow pulse is expected in this case. This pulse must contain odd
number of extremely narrow sub-pulses ($1, 3, 5$ or more with
smoothly falling intensities) (Fig.3,6). To be seen during
milliseconds such transient must have very short rotation period.

7. Li [15] put forward the model of a relic disk retained after
the supernova explosion or formed from the captured interstellar
matter. The neutron star operates as a propeller when the disk
penetrates inside the light cylinder and quenches the generation
of the of $e^{\pm}$ - plasma. An emission is switched on if disk
goes out  the light cylinder. In one case the braking is caused by
the magneto-dipole radiation, in the another one an additional
moment is taken away by the pulsar wind, and the pulsar is slowing
down faster.

    Wang et al. [16] detected IR radiation from the cold disk around the isolated young
X-ray pulsar 4U 0142+61. This is the first evidence of the
disk-like matter around the neutron star.

A surrounding plasma and a braking connected with it can explain
relatively long rotation periods comparing with normal radio
pulsars.

In this model the angle $\beta$ is expected to be high.

\section{Angles between rotation and magnetic axes}

As we declared in the previous section, the angle $\beta$ between
the rotation axis of the given object and its magnetic moment was
very important for the choosing of the adequate model. Here we
will discuss some possibilities of the evaluation of this angle
for a number of sources under consideration.

Using a schematic presentation of the radiation-cone geometry
(Fig.7), we can write three following equations [18]:

\begin{equation}
\begin{array}{lcr}
\sin \beta & = & C \sin (\zeta - \beta),\\
\cos \theta & = & \cos \zeta \cos \beta + D \sin \beta \sin \zeta,\\
\theta & = & n (\zeta - \beta)\\
\end{array}
\end{equation}

for three unknowns: the angle $\beta$ between the rotation axis
($\mathbf  \Omega$) and the magnetic moment ($\vec \mu$), the
angle $\zeta$ between the line of sight ($\mathbf L$) and $\mathbf
\Omega$, and the angular radius $\theta$ of the radiation cone,
assuming that it is connected to open magnetic-field lines.

\begin{figure}[h]
\epsfysize=10cm \epsfbox{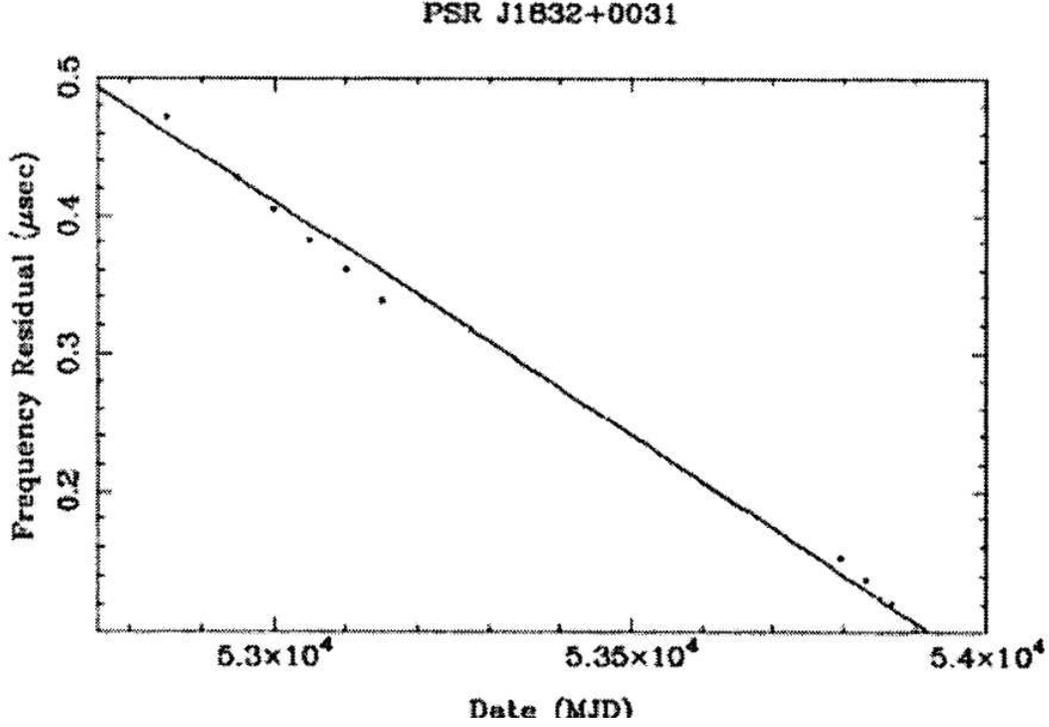}
   \caption{Variation of the rotational frequency of PSR J1832+0031.}
\end{figure}

   In (1) $n$ is the fraction of the angular radius of the cone where the line of sight passes and

\begin{equation}
C = _(d \psi / d \Phi)_{max} = \sin \beta / \sin(\zeta - \beta)
\end{equation}

is the maximum derivative of the linear-polarization position
angle in the average profile, which is obtained from the
relationship describing the position angle $\psi$ as a function of
longitude $\Phi$ (Fig. 7) [17]:

 \begin{equation}
\tan \psi = \frac{\sin \beta \sin \Phi}{\sin \zeta \cos \beta -
\cos \zeta \sin \beta \cos \Phi}
\end{equation}

The factor $D = \cos (W_{10}   / 2$) is determined by the observed
width $W_{10}$  of the mean profile at the $10\%$ level. The value
of n can be estimated from the shape of the mean profile.

Now we consider some pulsars with measured polarization
characteristics and known pulse profiles.

{\large\bf PSR B1931+24}

The radio pulsar PSR B1931+24 (J 1933+2421) emits pulses over five
to ten days, then sharply is switched off, remaining undetectable
over the next 25-30 days [19].

We have for this pulsar at a frequency close to 400 MHz [8,20]

  $$
C = 5.5, \, W_{10}  = 37.5^{\circ} , \, D = 0.95, \quad and \quad
n \approx 2.
$$

Thus, the problem of finding the angles $\beta, \, \zeta$, and
$\theta$ in this case is reduced to solving the system of
equations

 \begin{equation}
\begin{array}{lcr}
\sin \beta & = & 5.5 \sin (\zeta - \beta),\\
\cos \theta & = & \cos \zeta \cos \beta + 0.95 \sin \beta \sin \zeta,\\
\theta & = & 2 (\zeta - \beta)\\
\end{array}
\end{equation}

\begin{figure}[h]
\epsfysize=8cm \epsfbox{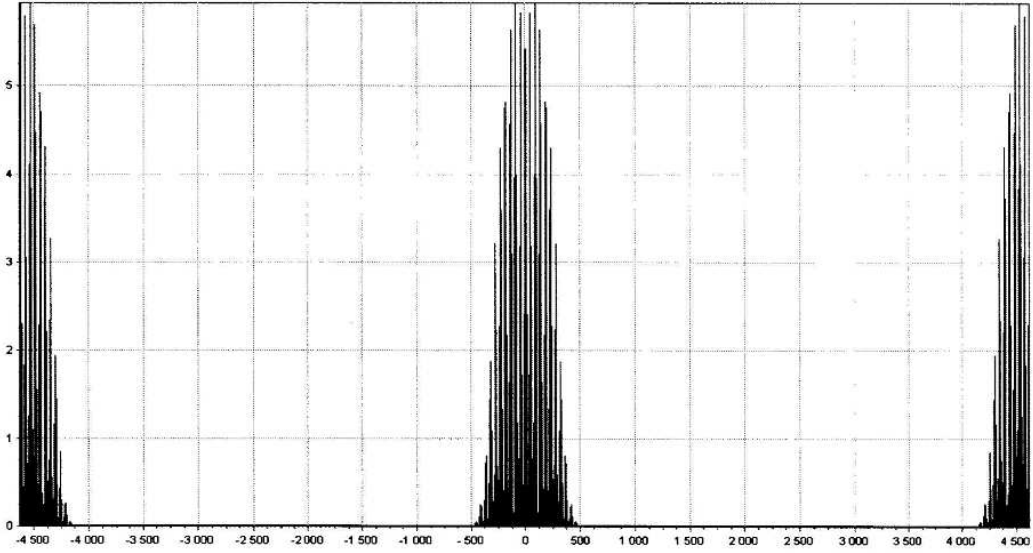}
   \caption{Simulated light curve of PSR J1819-1458 [14].}
\end{figure}

The system (4) can be transformed to the following form [18]:

\begin{equation}
b_3  y^ 3 + b_2 y^2  + b_1 y + b_0  = 0,
\end{equation}

\begin{equation}
\tan \beta = \frac{C (1 - y^2 )^{1/2}}{1 + Cy },
\end{equation}

where $y = \cos \zeta$ and coefficients $b_i$ are determined as

 \begin{equation}
\begin{array}{lcr}
b_3 & = & 2 C^3 (1 - D)^2 ,\\
                           b_2 & = & C^4 (1 - D)^2  + C^2 (D^2  - 6 D + 5) - 4,\\
                           b_1 & = & 2 C [C^2 (2 - D - D^2 ) - 2 - D],\\
                           b_0 & = & C^4 (1 - D^2 ) - C^2 (2 + D^2 ) + 1.\\
\end{array}
\end{equation}

\begin{figure}[h]
\epsfysize=10cm \epsfbox{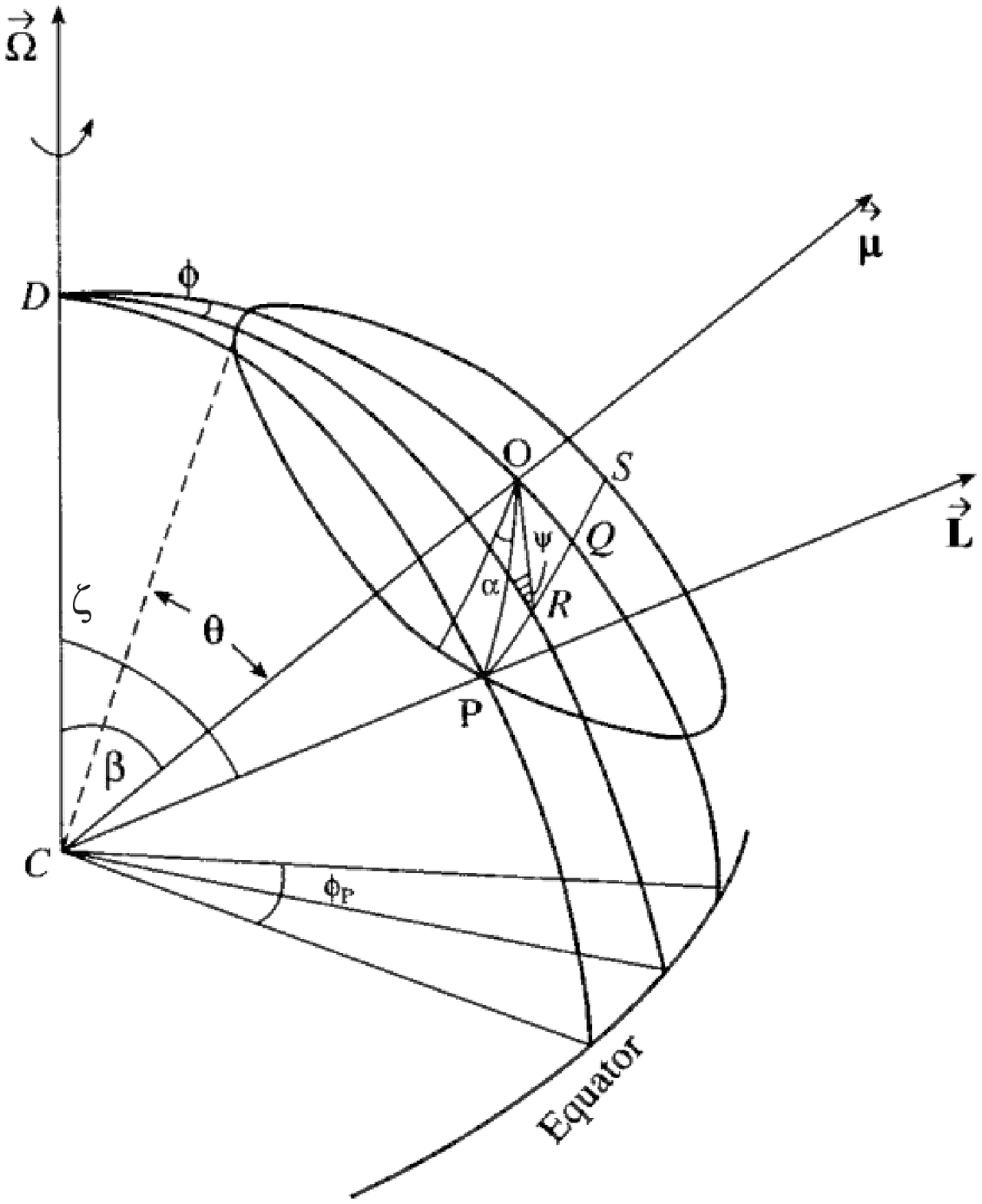}
   \caption{Geometry of the radiation cone in the polar cap model.}
\end{figure}

For the observed parameters we obtain the following cubic equation
for y:

  \begin{equation}
0.83 y^3  + 4.41 y^2  + 16.63 y + 2.42 = 0,
\end{equation}

which can be solved using the formula of Cardano [21]. This
equation has only one real-valued root, $y = - 0.15$, which
corresponds to $\zeta = 98^{\circ}.7$. We obtain from (6) the
value $\beta = 88^{\circ}.2$, and from the last equation of system
(4), $\theta = 21.0^{\circ}$.

The analysis of equations under consideration shows that the main
errors in the obtained solutions are caused by uncertainties in
the measured value of C.

The error $\Delta D = \sin (W_{10} / 2) \Delta W_{10} / 2$ is
rather small ($ < 10^{-3}$ ) and can be excluded from the
consideration. If we put $\Delta C = 0.1$, then for $C = 5.4$ the
angle $\beta = 76^{\circ}.1, \, \zeta = 86^{\circ}.5$ and for $C =
5.6$ $\beta = 93^{\circ}.7, \, \zeta = 110^{\circ}.0$. Hence, this
pulsar is nearly orthogonal rotator for $C = 5.5 \pm 0.1$.

The calculated  values of angles ($\zeta = 98^{\circ}.7$ and
$\beta = 88^{\circ}.2$) correspond to the pulsar geometry
presented in Fig.8. Analysis of this geometry leads to the
following conclusions.

1. Since PSR B1931+24 is an orthogonal rotator, pulses should be
observed from both its poles, i.e., the true period of this pulsar
should be twice the usual adopted value, and equals to $P = 1.626$
~s.

2. We expect different mean-profile shapes and integrated energies
for even and odd pulses, since the line of sight changes its
orientation with respect to the center of the radiation cone by
$3.6^{\circ}$  in the transition between the poles.

3. If the radiation cone is determined by open field lines  for
$\beta \approx 90^{\circ}$ , its angular radius should be [22]

  \begin{equation}
\theta = 0.54  (r / r_{LC})^{1/2},
\end{equation}

where $r$ is the distance from the center of the neutron star, and
$r_{LC}  = cP / 2 \pi$ is the radius of the light cylinder. For
the calculated value $\theta = 21.0^{\circ}$ , we find that the
radiation at a frequency of about 400 MHz is generated at a level
$r / r_{LC}  = 0.46$. For $P = 1.626$ ~s, the corresponding
distance is $r = 3.57 \times 10^9$ ~cm.

\begin{figure}[h]
\epsfysize=10cm \epsfbox{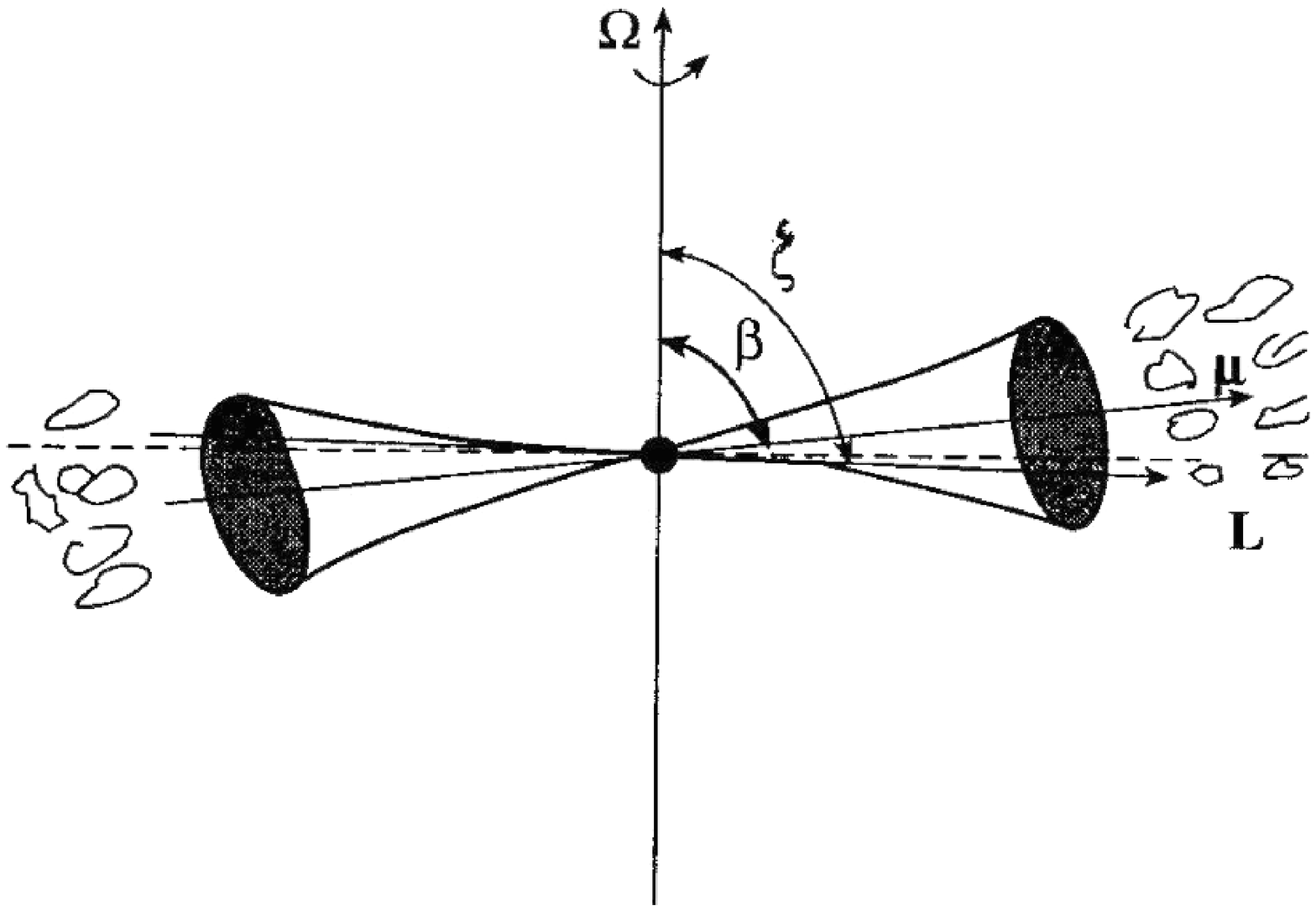}
   \caption{Radiation geometry for the pulsar PSR B1931+24.}
\end{figure}

As it was mentioned in the previous section, the possible cause of
the switching on and off of the radiation in PSR B1931+24 was the
presence of  a relic disk around the pulsar. In particular it can
precess with a  period of about 30-40 days. If such a disk is in
the equatorial region, it can occasionally "lock up" the pulsar
radiation. Searches for traces of such a disk, e.g., of its
infrared radiation, are necessary. The possible existence of relic
disks around pulsars with transient radiation was also considered
in [15,16]. A decrease in $dP / dt$ can also be related to the
presence of plasma of the relic disk at the times of switching
off. Partial or full blocking of the radiation inside the
magnetosphere decreases the loss of angular momentum of the
pulsar. Meanwhile, the escape of relativistic particles to the
surrounding plasma partially persists, and this pulsar wind
provides a spin-down in the periods of switching off of the
observed radiation. A small amount of disk accretion also can spin
up the neutron star, decreasing $dP / dt$ somewhat. Generally
speaking, the presence of precession is not a mandatory element in
the proposed model. Locking and unlocking of the radiation cone
could be due to inhomogeneity of the surrounding disk. In this
case, due to differential disk rotation, dense inhomogeneities
sometimes fall in the line of sight, hindering the propagation of
the pulsar radiation. Essentially, this type of situation can be
observed in, e.g., other Rotating RAdio Transients (RRATs).

Beskin and Nochrina [23] discussed the role of current losses in
the magnetosphere of PSR B1931+24. These losses could account for
the observed jump $(d\Omega / dt)_{on}  / (d\Omega / dt)_{off}  =
1.5$. In "off" periods, the magnetosphere is not filled with
plasma, and angular-momentum losses are due to magneto-dipole
radiation. When the radiation is observed, braking of the neutron
star is due to current losses; the ratio of these energy losses is
then

  \begin{equation}
\frac{ W_{md} }{ W_c } = \frac{B_s^2 R^6 \Omega^4 \sin^2 \beta}{6
c^3} : \frac{f_*^2 B_s^2 R^6 \Omega^4 \cos^2 \beta}{4 c^3} =
\frac{2 \tan^2 \beta }{3   f_*^2}
\end{equation}

However, using $\beta = 88^{\circ}.2$ and the dimensionless area
of the polar cap $f_* = 1.96$, we obtain

$$
(d\Omega / dt)_{on} / (d\Omega / dt)_{off}  = W_{md} / W_c = 264.
$$

Thus, magneto-dipole losses in the "on" state of PSR B1931+24
should be considerably greater than the current losses in the
"off" state.

{\large\bf PSR B0656+14}

Weltewrede et al. [6] proposed that the burst-like character of
the radiation of PSR B0656+14 is similar to the behavior of RRATs.
In this connection, it is interesting to estimate the angles
between its axes. We could use the system (1), but, in the case of
PSR B0656+14, it was sufficient to obtain simpler estimates based
on the profile shape and derivative of the polarization position
angle. For this pulsar, the maximum derivative at frequencies near
400 MHz is $C = 1.28$ [20]. For such a small value of $C$, we can
use the formula [2, 3]

    \begin{equation}
C \ge 3.24\ sin \beta,
\end{equation}

which yields $\beta < 23^{\circ}.3$. The conclusion that the angle
$\beta$ is small also follows from the simple (one  component)
shape of the pulsar pulse. Indeed, we have in this case [18]

    \begin{equation}
\zeta - \beta =3 \theta / 4
\end{equation}

On the other hand, it follows from Fig. 9 that

  \begin{equation}
\zeta - \beta = \theta \sin \alpha
\end{equation}

\begin{figure}[h]
\epsfysize=8cm \epsfbox{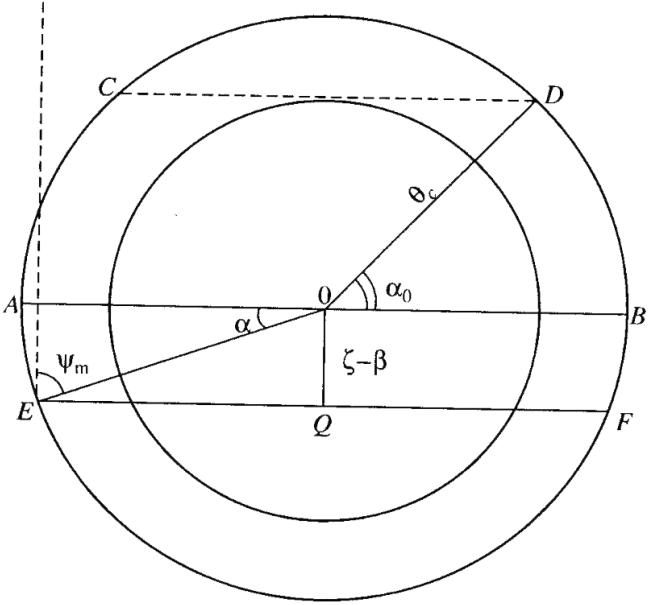}
   \caption{Cross section of the radiation cone.}
\end{figure}

As a result, for $W_{10}  = 35^{\circ}.3$ [20], we obtain $\theta
= 17^{\circ}.8, \, \zeta - \beta = 13^{\circ}.3$, and it follows
from the first equation of (1) that $\beta = 17^{\circ}.2$. These
estimates demonstrate that $\beta \sim 20^{\circ}$ in PSR
B0656+14, and this pulsar is a "normal" pulsar and not a
transient. The variations in its radiation (including outbursts)
have intrinsic origins, related to the particulars of its
radiation mechanism, not its specific geometry or the presence of
a relic disk.

{\large\bf PSR J 1810-197}

Kramer et al. [24] present polarization observations of the radio
emitting magnetar AXP J1810-197. Its emission is nearly $80-95\%$
polarized. The position angle swing has a low maximal slope $C \le
1$. Using the first equation from the system (1) or the inequality
(11), we conclude that $\beta < 20^{\circ}$. Hence, the main
suggestion of the drift model is fulfilled for this object. It is
nearly aligned rotator with the modulation of emission like
presented in Fig. 3. The  solution for all parameters by using
profiles from [24] is rather uncertain because we don't know a
real rotation period of this pulsar and can't calculate precisely
the parameters $D$ and $C$ to use the system (1).

\section{Conclusions and discussion}

1. As the example of the PSR J 1810-197 shows the drift model can
be used for the description of the peculiarities of AXP emission.
In this model real rotation periods must be much shorter than
observed intervals between the successive pulses. The smallness of
the angle $\beta$ provides very wide X-ray pulses but very narrow
radio pulses with extremely narrow sub-pulses.

2. The angles between the various axes in the pulsar PSR B1931+24
estimated from the observed profile shape and polarization data
indicate that this pulsar is an orthogonal rotator. The magnetic
moment is inclined to the rotation axis by the angle $\beta =
88^{\circ}.2$, while the angle between the line of sight and the
rotation axis is $\zeta = 98^{\circ}.7$.

3. The possible origin of the switching on and off of the observed
radio emission of the pulsar is precession of a relic disk with a
period of about 35 days.

4. For other pulsars with prolonged switching off of the observed
radiation (transients), polarization measurements are required to
enable estimation of the angles between their axes, and to test
whether they are orthogonal rotators.

If all transients prove to be orthogonal rotators, a common
picture for them could be as follows. A relic disk almost
continually screens the radiation of the neutron star from us,
but, owing to the disk's inhomogeneity, it can have gaps, through
which the pulses can "leak out" toward the observer during a
limited time.

To check the model proposed in this paper, it is extremely
important to search for manifestations of relic disks around
transients and periodicity in the variations of their radiation.
The results obtained in [16] inspire optimism. Transient objects
are fairly young ($\tau = P / (2 dP / dt) \sim 10^6$ years);
hopefully, the material that was ejected during the supernova
explosion has not yet completely dispersed in the interstellar
medium.

Probably there is the bimodality of anomalous pulsars. AXPs, SGRs
and some radio transients belong to the population of aligned
rotators with the angle between the rotation axis and the magnetic
moment $\beta < 20^{\circ}$. These objects are described by the
drift model, and their observed periods are connected with
periodicity of drift waves. Other sources have $\beta \sim
90^{\circ}$, and switchings on and switchings off of their
radiation are caused by accretion phenomena connected with a relic
(debris) disc surrounding them.

{\it Acknowledgements} This work was supported by the Russian
Foundation for Basic Research (project code 06-02-16888) and the
National Science Foundation (project 00-98685).

\end{document}